\documentclass[prd,aps,preprint,amsmath,nofootinbib,amssymb,eqsecnum]{revtex4-1}
\usepackage{subfig}
\usepackage{verbatim,graphics,graphicx,color,slashed}
\usepackage{hyperref}

\def\GeV{\textrm{GeV}}
\def\TeV{\textrm{TeV}}

\begin{document}

\title{\Large  Vacuum Stability in the Standard Model}

\author{Yong Tang\footnote{ytang@phys.cts.nthu.edu.tw}}
\affiliation{Physics Division, National Center for Theoretical Sciences, Hsinchu\\ 
Department of Physics, National Tsing Hua University, Hsinchu, Taiwan}

\date{\today}

\begin{abstract}
The long-awaited Higgs particle $H$ around 125 GeV has been observed at the LHC. Interpreting it as the standard model Higgs boson and if there is no new physics between electroweak and Planck scale, we then don't have a stable vacuum. Here, we give a brief review of the electroweak vacuum stability and some related theoretical issues in the standard model. Possible ways to save the stability are also discussed.  

\end{abstract}


\maketitle

\section{Introduction}
The long-awaited Higgs particle $H$ around 125 GeV has been observed at the LHC by ATLAS \cite{atlas:2012gk} and CMS \cite{cms:2012gu} with the combined significances of 5.9 and 5.0 standard deviations, respectively. Excesses of events have been shown in various channels, such as
$H\rightarrow \gamma\gamma$, $H\rightarrow ZZ^{*}$ and $H\rightarrow WW^{*}$. Although the precise determination of the new particle's properties requires future accumulation of data, the results at the moment are fully consistent with the standard model(SM). 

The Higgs field plays a special role in the SM. Not only it is the only scalar particle, but also it is responsible for all other particles' masses. Modern particle physics is tightly related with symmetry principles. Particles in the SM do not interact randomly, but follow a pattern described by gauge theory. The gauge group for the SM is the $ SU(3)_C\times SU(2)_L\times U(1)_Y$(${}_C$ stands for color, ${}_L$ for left-handed, and ${}_Y$ for hypercharge), of which $SU(3)_C$ rules the strong interaction and $SU(2)_L\times U(1)_Y$ for the electroweak interactions. And the Higgs field governs the breaking of $SU(2)_L\times U(1)_Y$ into $U(1)_{Q}$(${}_Q$ for electric charge).   

SM is ultra-violet(UV) complete and renormalizable due the Higgs mechanism, which means that it can be valid to very high energies. Although there exist experimental observations that motivate to extend the SM, they can be easily accommodated by slightly extension without spoiling the UV properties of SM. On the other hand, the intrinsic theoretical issues, such as hierarchy problem, could guide us to modify the theory completely.  

In the brief review, we are concerned of vacuum stability, an other theoretical issue in the SM. We first give an overview of the Higgs sector in the SM and then discuss several theoretical constraints on the mass of the Higgs boson. Later, vacuum stability, metastability and instability are briefly reviewed and finally the summary is given. Details and complete discussions are referred to \cite{Krasnikov, Sher:1988mj,Schrempp:1996fb}.

\section{Higgs Sector in the Standard Model}\label{sec:higgs}
The standard model composes of fermion, gauge and scalar sectors. The scalar particle is usually called Higgs boson, due to its role played in the spontaneous symmetry breaking of electroweak group, $SU(2)_L\times U(1)_Y$. In the SM, except for the neutrinos, all fermions are massive and have left-handed and right-handed parts, denoted collectively and respectively as $\psi_L$ and $\psi_R$, 
\begin{align}
L = 
\left( \begin{array}{c}
\nu_{L} \\ l_{L}
\end{array}\right), \; 
Q=
\left( \begin{array}{c}
u_{L} \\ d_{L}
\end{array}\right), 
l_R, u_R, d_R
\end{align}
Since $\psi_L$ and $\psi_R$ transform differently, the mass term $m_\psi\left(\bar{\psi}_L \psi_R+\bar{\psi}_R \psi_L\right)$ is not gauge invariant. We need to add new fields to restore the symmetry. In the SM a single scalar $SU(2)$ doublet is introduced for the gauge invariant interactions,
\begin{equation}
-\mathcal{L}_Y= y_l\bar{L}\Phi l_R+y_u\bar{Q}\tilde{\Phi}u_R +y_d\bar{Q}\Phi d_R,\;
\Phi=\left( \begin{array}{c}
\phi^{+} \\
\phi^{0}
\end{array}\right),\;
\tilde{\Phi}=i\tau_2\Phi^\ast.
\end{equation}
Flavor and color index are ignored for discussion in this paper. 

The potential with all gauge invariant and renormalizable terms is 
\begin{equation}
V_{\Phi}= - \mu^2 \Phi^{\dagger}\Phi + \lambda \left(\Phi^{\dagger}\Phi\right)^2 + V_0
\end{equation}
$V_0$ is an irrelevant constant for quantum field theory but may be very important for cosmology. So we should omit $V_0$ in the following discussion. The minus sign in front of $\mu^2$ is necessary for having a vacuum state, 
$
\langle \Phi \rangle = \frac{1}{\sqrt{2}}
\left( 0 \;\; v \right)^{\textrm{T}},
$
to trigger the EW symmetry breaking, $SU(2)_L\times U(1)_Y \rightarrow U(1)_Q$. We can expand the doublet as
\begin{equation}
\Phi=\left( \begin{array}{c}
 G^{+} \\
\dfrac{1}{\sqrt{2}}\left(v + H + i G^{0}\right)
\end{array}\right).
\end{equation}
Here $H$ is the physical Higgs field. $G^+$ and $G^0$ are the Goldstone modes, to be "eaten" by $W^{+}$ and $Z$, respectively. After the breaking fermions and gauge bosons get masses,
\begin{equation}
m_f=\dfrac{y_f}{\sqrt{2}}v;\;m_W=\dfrac{1}{2}gv;\;m_Z=\dfrac{1}{2}\sqrt{g^2+g'^2}v,
\end{equation}
where $g=0.6519$ and $g'=0.3575$ are the $SU(2)_L$ and $U(1)_Y$ gauge couplings defined at scale $m_Z$, respectively. Photon, relic of the symmetry breaking, is associated with the unbroken $U(1)_Q$ and is still massless. 

In the unitary gauge, before symmetry breaking, the scalar doublet can be written as 
\begin{equation}
\Phi=\frac{1}{\sqrt{2}}
\left( \begin{array}{c}
 0 \\
\varphi
\end{array}\right).
\end{equation}
In such a gauge, only physical fields appear in the theory, ghosts and Goldstone modes do not show from the very beginning. Then the potential turns to a function of the single variable,
\begin{equation}
V_\varphi=-\frac{1}{2}\mu^2 \varphi ^2 + \frac{1}{4}\lambda \varphi ^4. 
\end{equation}
After the $\varphi$ develops a vev, $\varphi=v+H$, all particles in the SM get masses and for the Higgs particle $H$,
\begin{equation}
m_H^2=2\lambda v^2,\; v^2=\frac{\mu ^2}{\lambda}.
\end{equation}
Because $v=246.22$ in SM is determined by measuring the mass of weak gauge boson, then the sole undetermined parameter is the mass of Higgs boson $m_H$ or equivalently the self-interaction coupling $\lambda$. If we interpret the recently-discovered particle as the SM Higgs boson, the $m_H\simeq 125\GeV$, we now has fixed all the parameters in the SM.

\section{Theoretical Constraints}
Before the LHC era, various theoretical considerations has already constrained the parameters in the Higgs potential, including unitarity, triviality and vacuum stability \cite{Hambye:1996wb}. Of them, the later two are related to the potential and renormalization group running while unitarity bound comes from the constraints on the scattering process of longitudinal weak gauge bosons. In this section, we shall only concentrate on triviality and vacuum stability.

The discussion in Sec.~(\ref{sec:higgs}) is based on the tree level potential and it is not straightforward to see how constraints can be put on. For a realistic and consistent consideration, the quantum effective potential $V_{\textrm{eff}}$ is needed. Quantum loop corrections will make the mass parameter and coupling dependent on the sliding scale $\Lambda$,
\begin{equation}
V_{\textrm{eff}}=-\frac{1}{2}\mu(\varphi)^2\varphi^2+\frac{1}{4}\lambda(\varphi)\varphi^4.
\end{equation}
Here we have set the scale $\Lambda$ to the field value $\varphi$. For discussion of vacuum stability the first term can be omitted as long as large field value $\varphi \gg v$. The $\lambda$ for the quartic term is running with the energy scale $\Lambda$
\begin{equation}
\Lambda\frac{d}{d\Lambda}=\beta_\lambda,
\end{equation}
and at one loop order $\beta_\lambda$ is given by
\begin{equation}\label{eq:rgeoneloop}
\beta_\lambda=
 \frac{1}{(4\pi)^2 } \left[ 24 \lambda ^2-6 y_t^4+\frac{3}{8} \left(2 g^4+\left(g^2+g'^2\right)^2\right)+\left(-9 g^2-3 g'^2+12 y_t^2\right) \lambda \right].
\end{equation}  
For later convenience, we point out the origins of the individual terms in Eq.~(\ref{eq:rgeoneloop}). $24 \lambda ^2$ comes from the Higgs self-interaction's contribution, $-6 y_t^4$ from the top quarks loop, $\dfrac{3}{8} \left(2 g^4+\left(g^2+g'^2\right)^2\right)$ from the gauge boson loop and the last term from higgs field renormalization. The relative sign between bosonic and fermionic contributions would dramatically affect the UV behavior of the theory.
 
\noindent \textit{Triviality}:
First let us consider an illumination that $\lambda$'s contribution is dominant so that we can neglect the fermion and gauge boson terms in Eq.~(\ref{eq:rgeoneloop}). For an input value of $\lambda(v)$ at the scale $v$, $\lambda(\Lambda)$ at $\Lambda$ is determined by
\begin{equation}
\lambda(\Lambda)=\dfrac{\lambda(v)}{1-\dfrac{24}{(4\pi)^2}\lambda(v)\ln\dfrac{\Lambda}{v}}.
\end{equation} 	
When the denominator vanishes, we get Landau pole at the scale $\Lambda_{\infty}$,
\begin{equation}
\Lambda_{\infty} = v \exp{\frac{2\pi^2}{3\lambda(v)}}=v \exp{\frac{4\pi^2 v^2}{3m^2_H}}
\simeq v\times\left(5\times 10^5\right)^{\frac{246^2}{m^2_H}}
\end{equation}
For typical values of $m_H$, we have $\Lambda_{\infty}\simeq 10^3 \GeV$ for $m_H=700 \GeV$, $\Lambda_{\infty}\simeq 10^8 \GeV$ for $m_H=246 \GeV$ and $\Lambda_{\infty}\simeq 10^{24}\GeV$ for $m_H=125 \GeV$. If we would like to avoid the Landau pole, the $\lambda$ at the input should be zero, leading to a non-interacting and trivial theory \cite{Lindner:1985uk} which can not provide the spontaneous symmetry breaking. However, practically it is still possible and safe to leave Landau pole beyond the Planck scale. All we need is a light Higgs boson. 

\begin{figure}[tb]
\centering
\includegraphics[scale=0.6]{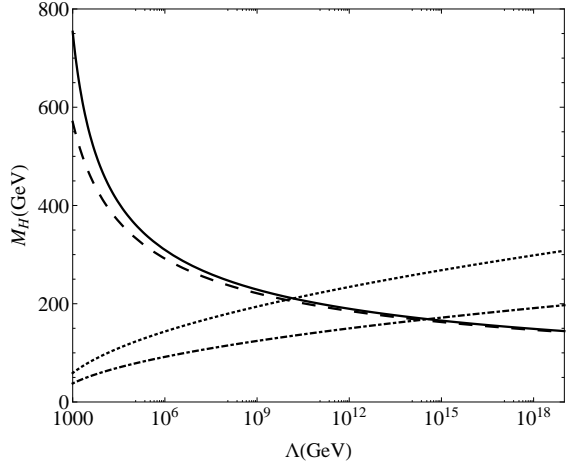}
\caption{An illustrating figure to show the perturbative and vacuum stability constraints. The solid and dashed lines corresponds $\Lambda_{\infty}$ and $\Lambda_{4\pi}$, respectively. And the dot-dashed and doted lines displays the vacuum stability constraints for $m_f=40 \GeV$ and $50 \GeV$, respectively. \label{fig:constraint}}
\end{figure} 

If we set $\lambda(\Lambda_{x})=x$, we have
\begin{equation}
\Lambda_{x}=v\exp{\left[\dfrac{4\pi^2}{3}\left(\frac{246^2}{m^2_H}-\frac{1}{2x}\right)\right]}.
\end{equation}
As long as $\Lambda$ is less than $\Lambda_x$, we have $\lambda(\Lambda)<x$. Perturbatively we can have a calculable theory up to energy scale up to $\Lambda_{4\pi}$. The fig.~(\ref{fig:constraint}) shows that how the scale at which the theory turns to non-perturbative depends on the $m_H$ defined at the electroweak scale. The two lines(solid and dashed) correspond $\Lambda_\infty$ and $\Lambda_{4\pi}$. For a light Higgs boson, SM can be perturbative all the way to Planck scale, and it gets non-perturbative near $\mathcal{O}(10\TeV)$ if $m_H>400\GeV$.

\noindent \textit{Vacuum Stability}:
As shown above, lighter Higgs boson means larger non-perturbative scale. However, if too light the theory is confronted with another problem, vacuum instability.	This is due to the heavy quark's contribution. Now we investigate the fermion's effect. Again just considering the only term in the $\beta_{\lambda}$
\begin{equation}
\beta_\lambda=
 \frac{1}{(4\pi)^2 } \left[-6 y_f^4 \right].
\end{equation} 
we can solve $\lambda(\Lambda)$ analytically when neglect the running for the $y_f$
\begin{equation}
\lambda(\Lambda)=\lambda(v)-6y^4_f\ln\frac{\lambda}{v}.
\end{equation}
For a complete and consistent investigation, we must solve all the coupled RGEs. But just for showing the physical effect of $y_t$ the simplification is enough. It is immediate to see that at some scale, $\lambda_{\Lambda}$ can cross zero and turn to negative. Then the potential is unbounded and the electroweak vacuum $\varphi=v$ may be unstable due to the quantum tunneling. Two lower lines in Fig.~\ref{fig:constraint} show how the scale at which $\lambda(\Lambda)$ cross zero depends on the mass of the Higgs boson. Different lines display effects of the fermion's mass. 

\begin{figure}[tb]
\centering
\includegraphics[scale=0.7]{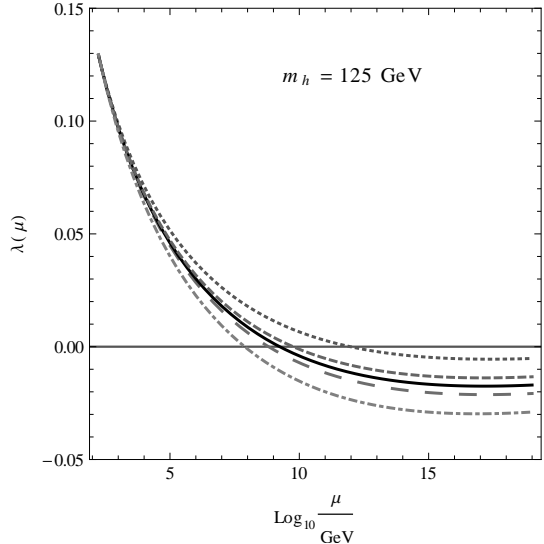}
\caption{Running of $\lambda(\mu)$ in standard model. Black solid line is plotted with $m_t=173.2$ GeV and $\alpha_s(M_Z)=0.1184$. Dotted and Dot-Dashed lines shows the effects of changing mass of $m_t$, for $m_t=171.4$ GeV and $m_t=175.0$, respectively. Two lines, closer to the solid one, display the effects of different value for $\alpha_s(M_Z)$, dashed one for $\alpha_s(M_Z)=0.1198$ and long dashed for $\alpha_s(M_Z)=0.1170$. }
\label{fig:SM_Stability}
\end{figure}
The realistic case in the standard model is show in the Fig.~(\ref{fig:SM_Stability}), where we show effects of variation on $M_t$ and $\alpha_s$. This is done by including one loop RGEs for top quark's Yukawa coupling,
\begin{equation}
\beta_{y_t} =
 \frac{y_t}{(4\pi)^2} \left[\frac{9}{2} y_t^2 -\frac{9}{4} g^2-\frac{17}{12}g'^2-8 g_s^2\right],
\end{equation}
and for the gauge couplings $g_i=\{g',g,g_s\}$,
\begin{equation}
\beta_{g_i} = \frac{1}{(4\pi)^2}b_i g_i^3,\; b=(41/6,-19/6,-7).
\end{equation}
Note that the above definition for $g_1$ is different from $g_{1}^{U}$ used usually for Unification consideration, $ g_1^{U}=\sqrt{\frac{5}{3}}g'$. In that case, the corresponding coefficients in the RGEs also need to change. The relevant input and RG equations in the numerical evaluations are listed below \cite{pdg},
\begin{align}
& m_H=125 \textrm{ GeV},\;m_t=173.2\pm 0.9 \textrm{ GeV},\;M_Z=91.188\textrm{ GeV},\nonumber \\
& \alpha_s(M_Z)=0.1184\pm 0.0007,\;\alpha(M_Z)=1/127.926, \;\textrm{sin}{}^2\theta(M_Z)=0.2312.
\end{align}

\section{Vacuum Stability and Effective Potential}
\begin{figure}[tb]
\centering
\subfloat[stable]{\includegraphics[scale=0.45]{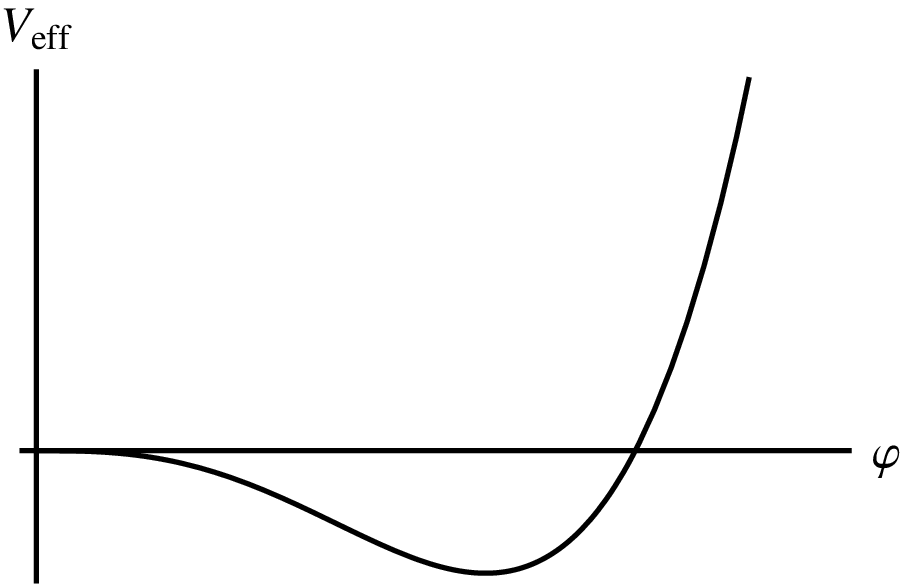}}
\subfloat[stable]{\includegraphics[scale=0.45]{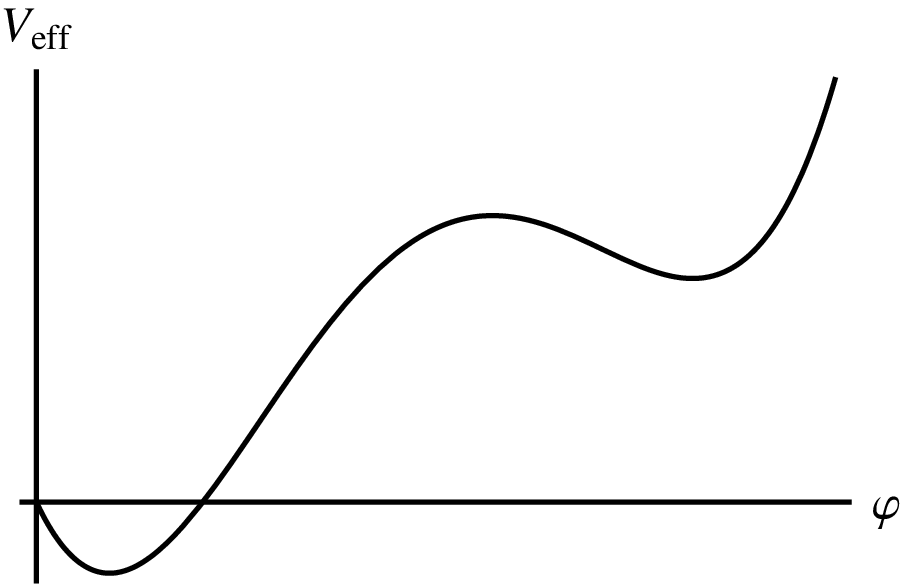}}
\subfloat[metastable]{\includegraphics[scale=0.45]{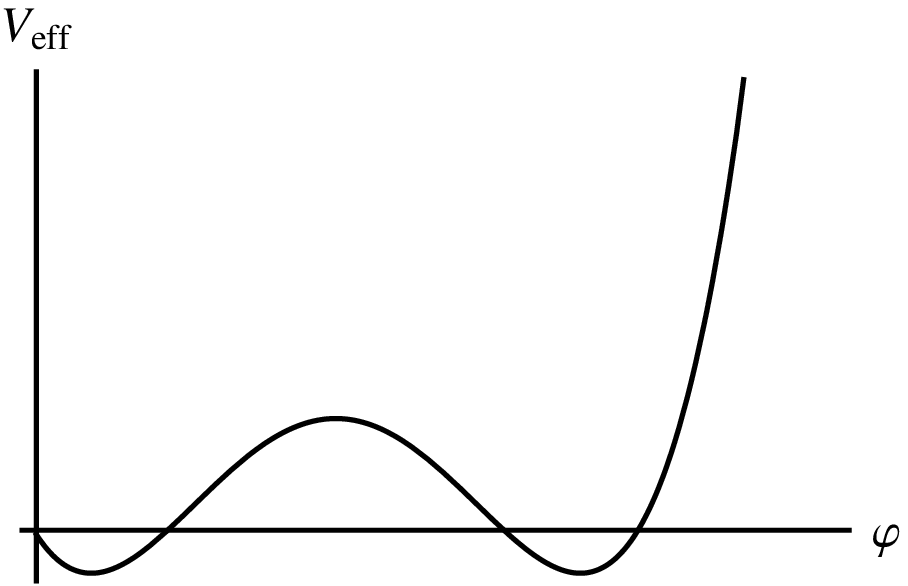}}\\
\subfloat[metastable]{\includegraphics[scale=0.45]{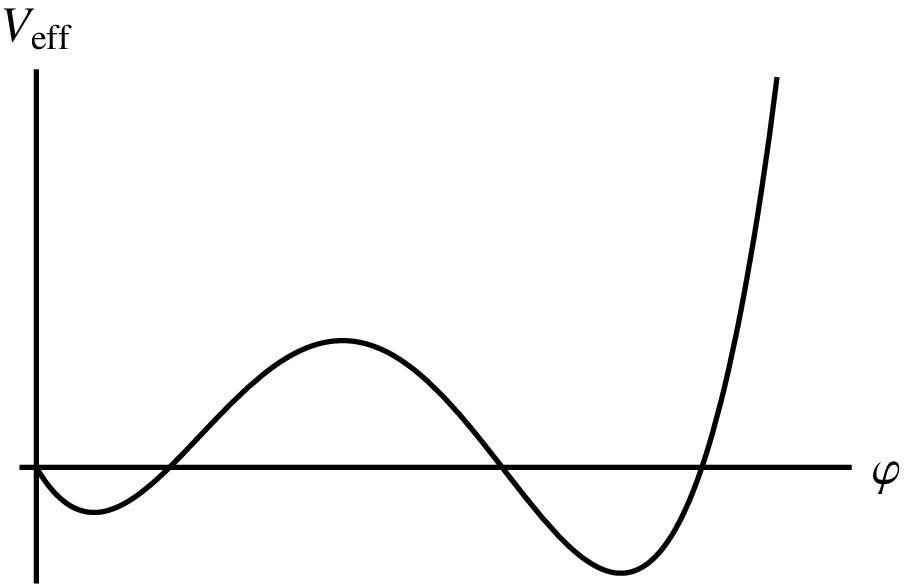}}
\subfloat[unstable]{\includegraphics[scale=0.45]{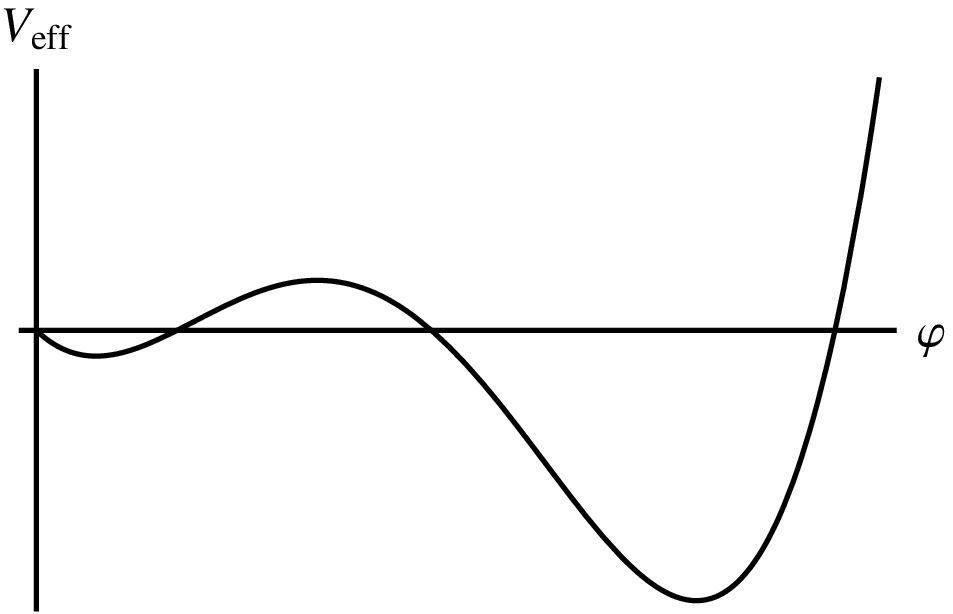}}
\subfloat[unstable]{\includegraphics[scale=0.45]{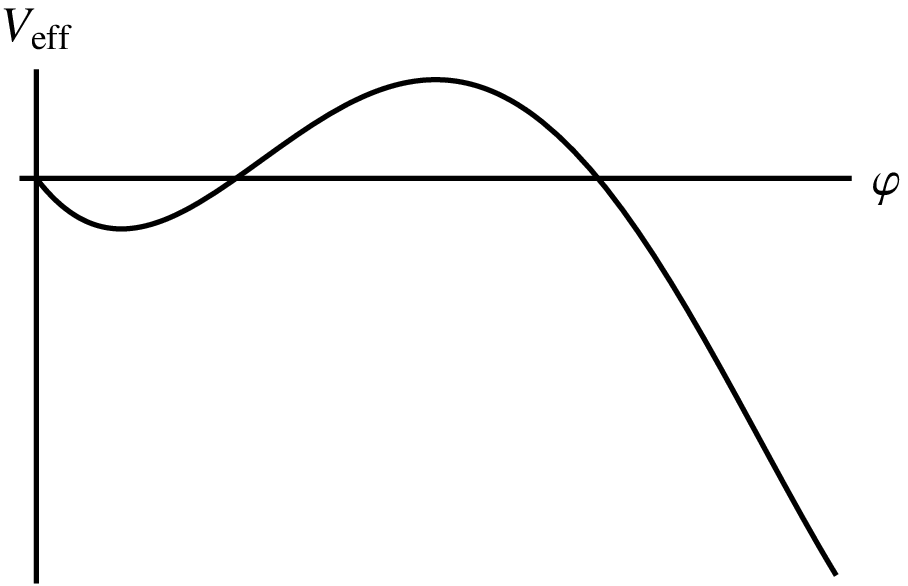}}
\caption{Various configurations of the effective potential. Local minimum near the original is the electroweak vacuum. \label{fig:Veff} }
\end{figure}	 

Details of vacuum stability \cite{Lindner:1988ww,Altarelli:1994rb,Casas:1996aq} depend on the behavior of the effective potential \cite{Ford:1992pn}. There are several configurations for the effective potential as shown in Fig.~(\ref{fig:Veff}). If the potential has only one minimum as case (a), then the vacuum is absolutely stable. Even there is another local minimum (b) but higher than the electroweak one, it is still stable and one can even use the configuration to inflate our universe in scalar-tenor framework \cite{Masina:2011un}. Other effects on inflation and fermion masses are discussed in \cite{Kobakhidze:2013tn,Xing:2011aa}. 
 
Vacuum stability up to Planck scale put constraint for the mass of the Higgs boson \cite{Holthausen:2011aa,Bezrukov:2012sa,Degrassi:2012ry},
\begin{equation}
m_H[\GeV]>129.5 + 1.4\left(\frac{m_t[\GeV]-173.1}{0.7}\right) - 0.5\left(\frac{\alpha_s(M_Z)-0.1184}{0.0007}\right).
\end{equation}
Complexity emerges when another local minimum at large field is as the same as or deeper than the EW one(c, d and e).  Then quantum tunneling effects from EW vacuum to the deeper one could make vacuum decay. If the life time is larger than the age of our universe, then the vacuum is metastable(c and d) \cite{Isidori:2001bm,Espinosa:2007qp}. If not, we have an unstable vacuum(e). The last one(c) the potential basically is unbounded from below, the vacuum then is definitely unstable. 

The formalism that quantitatively determines the decay rate of the false vacuum and the calculation procedures in general field theory was first developed semi-classically in \cite{Coleman:1977py} and quantum mechanically in \cite{Callan:1977pt}. The calculation in the SM is well described in \cite{Isidori:2001bm} and meta-stability in detail for $125\GeV$ SM Higgs boson is discussed in \cite{Degrassi:2012ry} .

\begin{figure}[hbt]
\centering
\includegraphics[scale=0.7]{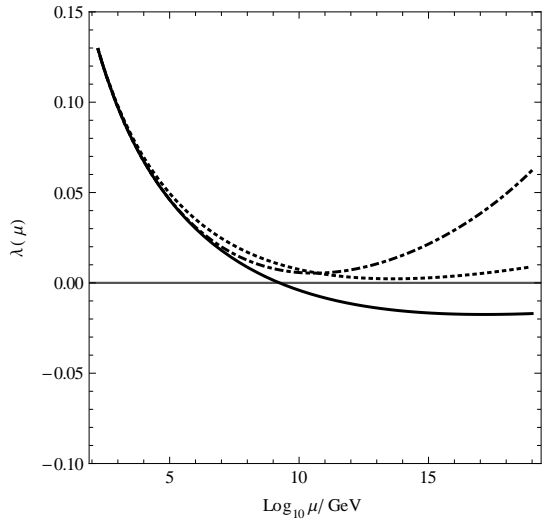}
\caption{Example models to save electroweak vacuum stable to Planck scale. The solid line is the standard model running. And the dashed line show a Higgs portal singlet dark matter's effect, which couplings to higgs boson $\lambda_{SH}$ and its self-coupling $\lambda_S$. Dot-dashed line shows a multiplet's contribution. \label{fig:savestability}}
\end{figure}

\noindent\textit{Save stability}:
There exists many and easy ways to save the electroweak vacuum stability up to Planck scale \cite{Chen:2012fa,EliasMiro:2012ay,Lebedev:2012zw,Rodejohann:2012px,Cheung:2012nb,Kannike:2012pe,Chao:2012mx,Allison:2012qn,Belanger:2012zr,Khan:2012zw,Patel:2012pi,Chao:2012xt,Dev:2013ff}. All of them are involved with changes on the standard model, based on different well-motivated considerations. For instance, adding a higgs portal singlet scalar will modifies the $\beta_{\lambda}$ with an extra bosonic contribution and this term can serve to stablize the vacuum to higher scale. One can also introduce many matter fields charged under SM gauge group and in this case gauge coupling could turn larger at higher scale, which indirectly affect $\beta_{\lambda}$ as extra bosonic contributions. Two examples are shown in Fig.~(\ref{fig:savestability}), dashed line for a singlet and dot-dashed one for a multiplet model, respectively. It is easily seen that $\lambda(\mu)$ is positive all the way to Planck scale, leading to a stable vacuum.

\section{Summary}
In this paper, we give a brief review of the theoretical constraints on the mass of the Higgs boson in the standard model. We mainly focus on triviality and vacuum stability. With the discovery of 125 GeV Higgs boson at the LHC, the standard model does not have a stable vacuum if no new physics is assumed. Ways to save vacuum stability to Planck scale are also briefly discussed and illustrated.

\end{document}